\documentclass[aps,prb,twocolumn,showpacs,preprintnumbers,amsmath,amssymb]{revtex4}
\usepackage{graphicx}% Include figure files
\usepackage{dcolumn}% Align table columns on decimal point
\usepackage{bm}% bold math

\def\TTOFA{Tb$_{1-x}$Th$_{x}$FeAsO }

\begin{document}

\title{Superconductivity above 50 K in Tb$_{1-x}$Th$_{x}$FeAsO}

\author{Lin-Jun Li, Yu-Ke Li, Zhi Ren, Yong-Kang Luo, Xiao Lin, Mi He, Qian Tao, Zeng-Wei Zhu, Guang-Han Cao\footnote[1]{Electronic address: ghcao@zju.edu.cn} and Zhu-An Xu\footnote[2]{Electronic address: zhuan@zju.edu.cn}}
\affiliation{Department of Physics, Zhejiang University, Hangzhou 310027, People's Republic of China}

\begin{abstract}
We have successfully synthesized one member of LnFeAsO (Ln stands
for lanthanides) with Ln=Tb. By partial substitution of Tb$^{3+}$
by Th$^{4+}$, superconductivity with onset $T_{c}$ up to 52 K was
observed. In the undoped parent compound, the magnetic moments of
Tb$^{3+}$ ions order antiferromagnetically (AFM) at $T_N$ of 2.5
K, and there exists an anomaly in the resistivity at $T^*$ of
about 124 K which corresponds to the structural phase transition
and/or AFM ordering of magnetic moments of Fe$^{2+}$ ions. This
anomaly is severely suppressed by Th doping, similar to the cases
of F doped LnFeAsO series. Thermopower measurements show that the
charge carrier is electron-like for both undoped and Th doped
compounds, and enhanced thermopower in Th doped samples was
observed.

\end{abstract}


%74.70.Dd Ternary, quaternary, and multinary compounds (including Chevrel phases, borocarbides, etc.)
%74.62.Dh Effects of crystal defects, doping and substitution
%74.62.Bf Effects of material synthesis, crystal structure, and chemical composition

\maketitle

The discovery of superconductivity at 26 K in
LaFeAsO$_{1-x}$F$_{x}$\cite{Kamihara08} has led to a type of high
temperature superconductors containing FeAs layers. The
fluorine-doped family includes CeFeAsO$_{1-x}$F$_{x}$ ($T_c$ = 41
K)\cite{Chen-Ce}, PrFeAsO$_{1-x}$F$_{x}$ ($T_c$ = 52
K)\cite{Ren-Pr}, NdFeAsO$_{1-x}$F$_{x}$ ($T_c$ = 52
K)\cite{Ren-Nd}, SmFeAsO$_{1-x}$F$_{x}$ ($T_c$ = 55K)
\cite{Chen-Sm,Chen-Sm2,Ren-Sm} and GdFeAsO$_{1-x}$F$_{x}$ ($T_c$ =
36 K)\cite{Chen-elements,Wen-Gd}, in addition to
LaFeAsO$_{1-x}$F$_{x}$. By using high-pressure synthesis method, a
family of oxygen-deficient superconductors $R$FeAsO$_{1-x}$
($R$=La, Ce, Pr, Nd and Sm) \cite{Ren-OD} and
GdFeAsO$_{1-x}$\cite{Ren-Gd} were discovered. Through the
Th$^{4+}$ substitution at the lanthanide site in GdFeAsO, we have
recently obtained Gd$_{1-x}$Th$_{x}$FeAsO superconductors with
$T_c$ up to 56 K.\cite{WangC}

The parent compounds of these superconductors have the general
formula LnFeAsO (Ln=lanthanide)~\cite{Quebe}, whose structure can
be described as the stacking of alternative
[Ln$_{2}$O$_{2}$]$^{2+}$ layers and [Fe$_{2}$As$_{2}$]$^{2-}$
layers. The [Ln$_{2}$O$_{2}$]$^{2+}$ layers act as a charge
reservoir while the [Fe$_{2}$As$_{2}$]$^{2-}$ layers are
superconductively active. Heterovalent substitutions of
F$^{-}$-for-O$^{2-}$ at the oxygen site or Th$^{4+}$-for-Ln$^{3+}$
at the lanthanide site in the [Ln$_{2}$O$_{2}$]$^{2+}$ layers
inject electrons onto [Fe$_{2}$As$_{2}$]$^{2-}$ layers, which
results in the occurrence of superconductivity. Unfortunately, the
parent compounds LnFeAsO were limited to Ln=La, Ce, Pr, Nd, Sm and
Gd in the literatures. The LnFeAsO family member with Ln=Tb was
mentioned as ternary 'TbFeAs'\cite{Terbuchte}, but it was not
reproduced as quaternary arsenide oxides later.\cite{Quebe} Very
recently a synthesis of TbFeAsO by high pressure reaction has been
reported and superconductivity with $T_{c}$ of 46 K has been
observed in F-doped TbFeAsO.\cite{TbF} However, there are few
reports on the physical properties of this member probably due to
the difficulty in obtaining single phase samples with high purity.
In this paper, we report the synthesis of highly pure single phase
TbFeAsO and the occurrence of superconductivity with $T_{c}$ as
high as 52 K induced by the partial substitution of Tb$^{3+}$ by
Th$^{4+}$. The magnetic properties of undoped TbFeAsO were also
investigated. Th doping also induces enhanced thermopower in the
normal state, which implies strong electron correlation in this
system.

Polycrystalline samples of \TTOFA ($x$=0, 0.1, 0.2) were
synthesized by solid state reaction in an evacuated quartz tube.
The starting materials are Tb, Tb$_{4}$O$_{7}$, ThO$_{2}$, Fe and
As, which are all with high purity ($\geq$ 99.95\%). First, TbAs
was presynthesized by reacting Tb tapes with As powders at 853 K
for 10 hours and then 1173 K for 15 hours. Similarly, FeAs was
prepared by reacting Fe with As powders at 853 K for 6 hours and
then 1030 K for 12 hours. Then, the powders of TbAs,
Tb$_{4}$O$_{7}$, ThO$_{2}$, Fe and FeAs were weighed according to
the stoichiometric ratio of \TTOFA. The weighed powders were mixed
thoroughly by grinding, and pressed into pellets under a pressure
of 200 MPa in an argon-filled glove box. The pressed pellets were
wrapped with Ta foil, and sealed in an evacuated quartz ampoule.
The sealed ampoule was slowly heated to 1453 K and kept for 48
hours. Finally the samples were furnace-cooled to room
temperature. During the sample preparation, special measures were
taken to avoid possible exposure to the fine particles of
Th-contaminated materials in the air.

Powder X-ray diffraction (XRD) was performed at room temperature
using a D/Max-rA diffractometer with Cu-K$_{\alpha}$ radiation and
a graphite monochromator. The XRD diffractometer system was
calibrated using standard Si powders. Lattice parameters were
refined by a least-squares fit using at least 20 XRD peaks. The
electrical resistivity was measured with a standard four-probe
method. The temperature dependence of dc magnetization was
measured on a Quantum Design Magnetic Property Measurement System
(MPMS-5). The applied field was 1000 Oe for the undoped parent
TbFeAsO sample and 10 Oe for the superconducting samples.

Fig. 1 shows the XRD patterns of the \TTOFA ($x$ = 0, 0.1, 0.2)
samples. The XRD peaks of the parent compound ($x$ = 0) can be
well indexed based on the tetragonal ZrCuSiAs-type structure with
the space group P4/$nmm$, and no obvious impurity phases were
detected. The refined lattice parameters are $a$ = 3.8994(3) {\AA}
and $c$ = 8.4029(5) {\AA} for $x$ = 0, smaller than those of the
neighbor member GdFeAsO in the LnFeAsO family \cite{WangC}. In the
case of the Th-doped samples, small amount of ThO$_2$ impurity can
be observed. The refined lattice constants are $a$ = 3.8998(3)
{\AA}, $c$ = 8.4090(6) {\AA} and $a$ = 3.9025(3) {\AA}, $c$ =
8.4131(6) {\AA} for $x$ = 0.1 and 0.2, respectively. Therefore,
the lattice parameters increases monotonically with increasing
$x$. Because the ionic size of Th$^{4+}$ is larger than that of
Tb$^{3+}$\cite{Shannon}, the above result suggests the successful
substitution of Tb$^{3+}$ by Th$^{4+}$. The incorporation of
relatively large Th$^{4+}$ ions can also relax the lattice
mismatch between Ln$_2$O$_2$ fluorite-type block layers and
Fe$_2$As$_2$ conducting layers.

\begin{figure}
\includegraphics[width=8cm]{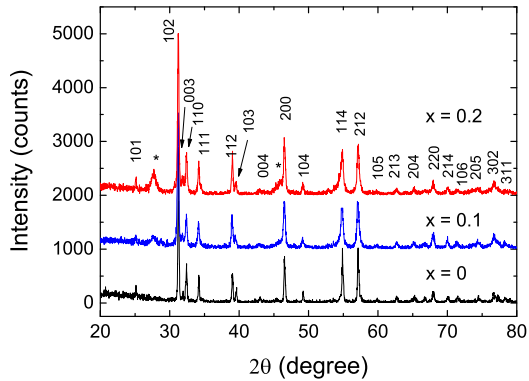}
\caption {(Color Online) Powder XRD patterns of \TTOFA ($x$=0,
0.1, 0.2) samples. The asterisks mark the impurities from
ThO$_{2}$.} \label{fig1}
\end{figure}

The temperature dependence of the resistivity for the undoped
parent compound ($x$=0) is shown in the upper panel of Fig. 2, and
the magnetic susceptibility is also shown in the lower panel.
Obviously the resistivity of the parent compound undergoes a sharp
decrease below $T^{*}$ of 124 K, where $T^*$ is defined as the
peak position in the curve of the derivative of resistivity versus
temperature (shown in the inset of upper panel). Such an anomaly
in the resistivity was also observed in the other LnFeAsO parent
compounds\cite{Kamihara08,wnl,Chen-Sm2}. Low temperature neutron
and X-ray diffraction studies\cite{Cruz,McGuire,Nomura} have
demonstrated that this anomaly is associated with a structural
transition and/or antiferromagnetic (AFM) transition in LaFeAsO.
However, $T^*$ is a little lower compared to the neighbor member
GdOFeAs \cite{WangC}. We suggest that the decrease in $T^*$ could
be correlated to the decrease in the radius of Ln$^{3+}$ ions in
the LnOFeAs parent compounds. The magnetic susceptibility shows
good Curie-Weiss behavior and a drop in susceptibility appears
around 2.5 K. Such a drop should be ascribed to AFM ordering of
the magnetic moments of Tb$^{3+}$ ions. It should be noted that
there is a decreasing tendency in $T_N$ as Ln goes from the light
rare earth element ($i.e$. Sm) to heavy one\cite{SmAFM,WangC}. The
effective magnetic moment is about 9.74 $\mu_B$ estimated from the
fitting parameters with the Curie-Weiss formula, which is close to
the magnetic moment (9.72 $\mu_B$) of free Tb$^{3+}$ ions.

\begin{figure}
\includegraphics[width=8cm]{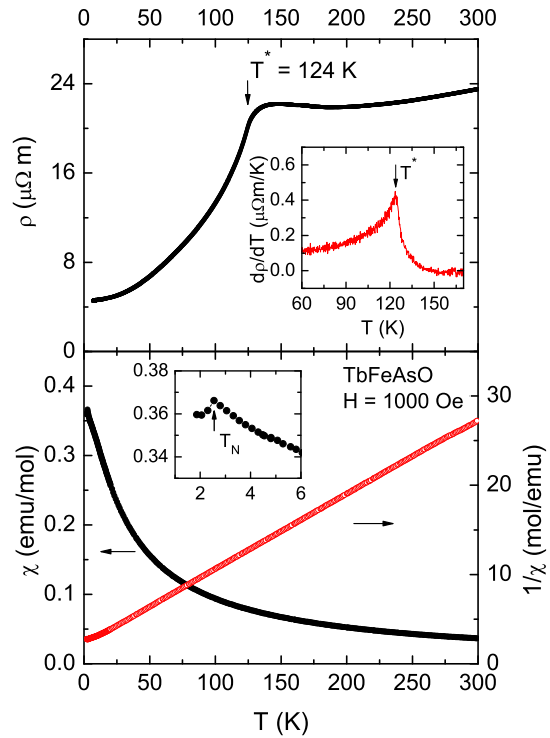}
\caption {(Color Online) Upper Panel: Temperature dependence of
electrical resistivity for undoped TbFeAsO samples. The inset
shows the derivative of resistivity as a function of temperature;
Lower Panel: Temperature dependence of the magnetic susceptibility
measured under $H$ = 1000 Oe. The inverse of the magnetic
susceptibility versus temperature is also shown to indicate the
Curie-Weiss behavior. The inset shows the enlarged plot at low
temperatures.}
\end{figure}

The change of resistivity with Th doping is shown in Fig. 3. Upon
Th doping, the resistivity anomaly around $T^*$ becomes less
prominent and superconductivity occurs. The onset transition
temperatures $T_c$ (defined as the onset point in the resistive
transition) are 45 K and 52 K for $x$ = 0.1 and 0.2, respectively.
Similar to the F-doped superconductors like LaFeAsO$_{1-x}$F$_x$,
there is a kink in the resistivity above $T_c$ in the "underdoped"
region. Even for the $x$ = 0.2 case, a kink is still observed
around 120 K. The inset of Fig. 3 shows the dc magnetic
susceptibility ($\chi$) versus temperature for the $x$ = 0.2
sample measured under zero-field cooling (ZFC) condition and $H$ =
10 Oe. A clear superconducting transition begins at 52 K,
consistent with the resistivity measurement. The magnetic
shielding fraction according to the ZFC susceptibility is above 50
\% at $T$ = 4 K. This implies bulk superconductivity induced by Th
doping. The reduced magnetic shielding fraction is often observed
in the poly-crystalline samples\cite{Kamihara08}, and it could be
due to the granular property of the samples and existence of a few
insulating impurities. The inhomogeneity of Th doping in the
samples could also broad the superconducting transition.

\begin{figure}
\includegraphics[width=8cm]{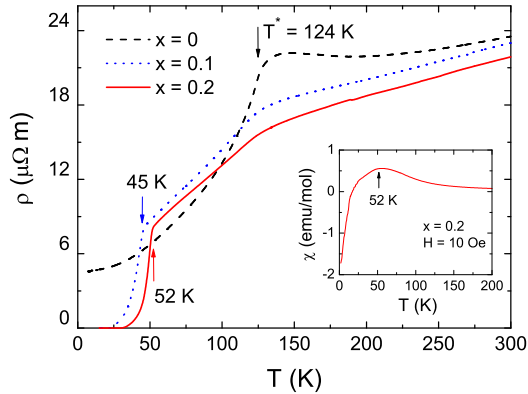}
\caption {(Color online) Temperature dependence of resistivity for
Tb$_{1-x}$Th$_{x}$FeAsO ($x$ = 0, 0.1, 0.2) samples. The inset
shows the magnetic susceptibility measured under
zero-field-cooling condition and $H$ = 10 Oe.  Note that there
exists a large contribution from the paramagnetic moments of
Tb$^{3+}$ in the normal state.}
\end{figure}

Fig. 4 shows the temperature dependence of thermopower for the
Tb$_{1-x}$Th$_{x}$FeAsO ($x$ = 0, 0.1, 0.2) samples. All of the
thermopower are negative at room temperature, which means that the
electron-like charge carriers dominate. For the undoped parent
compound, thermopower starts to increase abnormally from negative
to positive below $T^*$. Similar anomalous increase in the
thermopower below $T^*$ is also observed in the undoped LaFeAsO
\cite{McGuire}. This implies that the electronic state undergoes a
severe change after the structural phase transition. Such an
anomaly is completely suppressed in the Th doped samples. The
thermopower drops to zero quickly below the superconducting
transition temperature. The fact that both the anomaly in the
resistivity and that in the thermopower are severely suppressed by
Th doping indicates that superconductivity occurs as the AFM order
is suppressed, which implies that superconductivity and AFM order
are two competing ground states in this system. To our surprise,
the absolute value of thermopower, $|S|$, increases quickly with
Th doping, and the maximum in $|S|$ is about 108 $\mu$V/K for $x$
= 0.2, which is even comparable to that of the famous cobaltate
Na$_x$CoO$_2$ \cite{NCO}. It has been proposed that the doped
iron-oxypnictides can be promising thermoelectric materials in
refrigeration applications around liquid nitrogen temperatures
\cite{Dragoe}. A rough estimate of $|S|$ according to the Mott
expression gives a value of less than 10 $\mu$V/K for F doped
LnFeAsO \cite{Sefat}. More than 10 times enhanced value of $|S|$
strongly suggest that the electron correlation is very important
in the Fe-based oxypnictide superconductors.

\begin{figure}
\includegraphics[width=8cm]{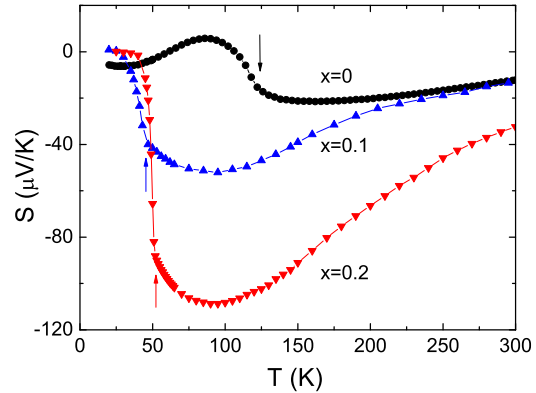}
\caption {(Color online) Temperature dependence of thermopower
($S$) for Tb$_{1-x}$Th$_{x}$FeAsO ($x$ = 0, 0.1, 0.2) samples. The
arrows indicate the position of $T^*$ for $x$ =0, and the
positions of $T_c$ for $x$ = 0.1, 0.2, respectively.}
\end{figure}

In summary, we have successfully synthesized one member of LnFeAsO
family, TbFeOAs, without applying high pressure. By partial
substitution of Tb$^{3+}$ by Th$^{4+}$, bulk superconductivity
with onset $T_{c}$ up to 52 K is induced. The magnetic moments of
Tb$^{3+}$ ions antiferromagnetically (AFM) order at $T_N$ of 2.5 K
in the parent compound. The anomaly in the resistivity
corresponding to the structural phase transition and/or AFM
ordering of magnetic moments of Fe$^{2+}$ ions is suppressed by Th
doping, which means that superconducting order and AFM order are
competing orders in this system. Furthermore, the absolute value
of thermopower increases abnormally with Th doping, implying
strong correlation between electrons in this sytem.

\begin{acknowledgments}
We would like to thank X. H. Chen, F. C. Zhang, and Y. Liu for
helpful discussions. This work is supported by the National
Scientific Foundation of China, the National Basic Research
Program of China (No.2006CB601003 and 2007CB925001) and the PCSIRT
of the Ministry of Education of China (IRT0754).
\end{acknowledgments}


\newpage
\begin{thebibliography}{0}
\bibitem{Kamihara08}Y. Kamihara, T. Watanabe, M. Hirano, and H. Hosono, J. Am. Chem. Soc. \textbf{130}, 3296 (2008).
\bibitem{Chen-Ce}G. F. Chen, Z. Li, D. Wu, G. Li, W.Z. Hu, J. Dong, P. Zheng, J.L. Luo, and N.L. Wang, Phys. Rev. Lett. \textbf{100}, 247002 (2008).
\bibitem{Ren-Pr}Z. A. Ren, J. Yang, W. Lu, W. Yi, G. C. Che, X. L. Dong, L. L. Sun, and Z. X. Zhao, Mater. Res. Inno. \textbf{12}, 1 (2008).
\bibitem{Ren-Nd}Z. A. Ren, J. Yang, W. Lu, W. Yi, X. L. Shen, Z. C. Li, G. C. Che, X. L. Dong, L. L. Sun, F. Zhou, and Z. X. Zhao, Europhys. Lett. \textbf{82}, 57002 (2008).
\bibitem{Chen-Sm}X. H. Chen, T. Wu, G. Wu, R. H. Liu, H. Chen, and D. F. Fang, Nature \textbf{354}, 761 (2008).
\bibitem{Chen-Sm2}R. H. Liu, G. Wu, T. Wu, D. F. Fang, H. Chen, S. Y. Li, K. Liu, Y. L. Xie, X. F. Wang, R. L. Yang, L. Ding, C. He, D. L. Feng, and X. H. Chen, Phys. Rev. Lett. \textbf{101}, 087001 (2008).
\bibitem{Ren-Sm}Z. A. Ren, W. Lu, J. Yang, W. Yi, X. L. Shen, Z. C. Li, G. C. Che, X. L. Dong, L. L. Sun, F. Zhou, and Z. X. Zhao, Chin. Phys. Lett. \textbf{25}, 2215 (2008).
\bibitem{Chen-elements}G. F. Chen, Z. Li, D. Wu, J. Dong, G. Li, W. Z. Hu, P. Zheng, J. L. Luo, and N. L. Wang, Chin. Phys. Lett. \textbf{25}, 2235 (2008).
\bibitem{Wen-Gd}P. Cheng, L. Fang, H. Yang, X. Y. Zhu, G. Mu, H. Q. Luo, Z. S. Wang, and H. H. Wen, Science in China G \textbf{51}, 719 (2008).
\bibitem{Ren-OD}Z. A. Ren, G. C. Che, X. L. Dong, J. Yang, W. Lu, W. Yi, X. L. Shen, Z. C. Li, L. L. Sun, F. Zhou, and Z. X. Zhao, Europhys. Lett. \textbf{83}, 17002 (2008).
\bibitem{Ren-Gd}J. Yang , Z. C. Li, W. Lu, W. Yi, X. L. Shen, Z. A. Ren, G. C. Che, X. L. Dong, L. L. Sun, F. Zhou, and Z. X. Zhao, Supercond. Sci. Technol. \textbf{21}, 082001 (2008).
\bibitem{WangC}C. Wang, L. J. Li, S. Chi, Z. W. Zhu, Z. Ren, Y. K. Li, Y. T. Wang, X. Lin, Y. K. Luo, S. Jiang, X. F. Xu, G. H. Cao, and Z. A. Xu, Europhys.
Lett. \textbf{83}, 67006 (2008).
\bibitem{Quebe}P. Quebe, L. J. Terbuchte, and W. Jeitschko, J. Alloys Compd. \textbf{302}, 70 (2000).
\bibitem{TbF}J. W. G. Bos, G. B. S. Penny, J. A. Rodgers, D. A. Sokolov, A. D. Huxley, and J. P. Attfield, Chem. Commun. \textbf{2008}, 3634 (2008).
\bibitem{Terbuchte}L.J. Terbuchte and W. Jeitschko, Z. Kristallogr. \textbf{186}, 291 (1989).
\bibitem{Shannon}R. D. Shannon, Acta Crystallogr. Sect. A \textbf{32}, 751 (1976).
\bibitem{wnl}J. Dong, H. J. Zhang, G. Xu, Z. Li, G. Li, W. Z. Hu, D. Wu, G. F. Chen, X. Dai, J. L. Luo, Z. Fang, and N. L. Wang, Europhys. Lett. \textbf{83}, 27006 (2008).
\bibitem{Cruz} C. de la Cruz, Q. Huang, J. W. Lynn, J. Li, W. Ratcliff II, H. A. Mook, G. F. Chen, J. L. Luo, N. L. Wang, and P. C. Dai, Nature \textbf{453}, 899 (2008).
\bibitem{McGuire}M. A. McGuire, A. D. Christianson, A. S. Sefat, B. C. Sales, M. D. Lumsden, R. Jin, E. A. Payzant, D. Mandrus, Y. Luan, V. Keppens, V. Varadarajan, J. W. Brill, R. P. Hermann, M. T. Sougrati, F. Grandjean, and G. J.
Long, arXiv:0806.3878.
\bibitem{Nomura}T. Nomura, S. W. Kim, Y. Kamihara, M. Hirano, P. V. Sushko, K. Kato, M. Takata, A. L. Shluger, and H. Hosono, arXiv:0804.3569.
\bibitem{SmAFM} L. Ding, C. He, J. K. Dong, T. Wu, R. H. Liu, X. H. Chen, and S. Y. Li, Phys Rev. B  \textbf{77}, 180510(R) (2008).
\bibitem{NCO} I. Terasaki, Y. Sasago, and K. Uchinokura, Phys.
Rev. B \textbf{56}, R12685 (1997).
\bibitem{Dragoe} L. Pinsard-Gaudart, D. B¨¦rardan, J. Bobroff, and N.
Dragoe, Phys. Stat. Sol. (RRL) \textbf{2}, 185 (2008).
\bibitem{Sefat} A. S. Sefat, M. A. McGuire, B. C. Sales, R. Jin, J. Y. Howe, and
D. Mandrus, Phys. Rev. B \textbf{77}, 174503 (2008).

\end{thebibliography}
\end{document}